\documentclass[12pt,preprint]{AwadSh_aastex}

\usepackage{graphicx}	
\usepackage{amsmath}	
\usepackage{amssymb}	
\usepackage{longtable}
\usepackage{rotate}
\usepackage{lscape}
\usepackage{latexsym,verbatim}

\slugcomment{Accepted by Astrophysics \& Space Science, 14 Mar 2017}

\shorttitle{Chemistry of NGC 2264 CMM3}
\shortauthors{Awad \& Shalabeia}

\begin{document}


\title{On the Chemistry of the Young Massive Protostellar core NGC 2264 CMM3}


\author{Z. Awad \and O. M. Shalabeia}
\affil{Astronomy, Space Science and Meteorology Department, Faculty of Science, Cairo University, Giza, Egypt} 
\email{zma@sci.cu.edu.eg}





\begin{abstract}
We present the first gas-grain astrochemical model of the NGC 2264 CMM3 protostellar core. 
The chemical evolution of the core is affected by changing its physical parameters such as the total density and the amount of gas-depletion onto grain surfaces as well as the cosmic ray ionisation rate, $\zeta$. We estimated $\zeta_{\text {CMM3}}$ = 1.6 $\times$ 10$^{-17}$ s$^{-1}$. This value is 1.3 times higher than the standard CR ionisation rate, $\zeta_{\text {ISM}}$ = 1.3 $\times$ 10$^{-17}$ s$^{-1}$. 
Species response differently to changes into the core physical conditions, but they are more sensitive to changes in the depletion percentage and CR ionisation rate than to variations in the core density. Gas-phase models highlighted the importance of surface reactions as factories of large molecules and showed that for sulphur bearing species depletion is important to reproduce observations. 

Comparing the results of the reference model with the most recent millimeter observations of the NGC 2264 CMM3 core showed that our model is capable of reproducing the observed abundances of most of the species during early stages ($\le$ 3$\times$10$^4$ yrs) of their chemical evolution. Models with variations in the core density between 1 - 20 $\times$ 10$^6$ cm$^{-3}$ are also in good agreement with observations during the early time interval 1 $\times$ 10$^4 <$ t (yr) $<$ 5 $\times$ 10$^4$. 
In addition, models with higher CR ionisation rates (5 - 10) $\times \zeta_{\text {ISM}}$ 
are often overestimating the fractional abundances of the species. However, models with $\zeta_{\text {CMM3}}$ = 5 $\zeta_{\text {ISM}}$ may best fit observations at times $\sim$ 2 $\times$ 10$^4$ yrs. 
Our results suggest that CMM3 is (1 - 5) $\times$ 10$^4$ yrs old. Therefore, the core is chemically young and it may host a Class 0 object as suggested by previous studies. 
\end{abstract}


\section{Introduction}
\label{intro}
NGC 2264 is a massive cluster-forming region where two main clumps are identified NGC 2264 -C and -D. The region is located at a distance 
of 760 - 913 pc and several studies adopted a distance of 800 pc (e.g. \citealt{sar11, wat15}). The NGC 2264 -C hosts 13 millimeter sources 
identified by single-dish and interferometric observations of the region \citep{war20,per06,per07,mau09}. These cores have typical radii 
of 0.04 pc, masses $\sim$ 2 - 40 M$_{\odot}$, and column densities of 0.4 - 6.0 $\times$ 10$^{22}$ cm$^{-2}$. The brightest IR source in NGC 2264-C 
is IRS1 which is a 9.5 M$_{\odot}$ B2 ZAMS\footnote{ZAMS stands for Zero Age Main Sequence} star known also as Allen object \citep{alen72}. 
From millimeter (mm) and submillimetre (sub-mm) continuum observations of NGC 2264, \citet{war20} estimated the size of the millimeter 
source CMM3 (called MM3 in their work) to be 0.1 pc $\times$ 0.05 pc and the mass to be 48 M$_{\odot}$ which will evolve into a 8 M$_{\odot}$ star. 
Their observations also revealed that the region has a visual extinction, A$_v$, range from 200 to 1600 magnitudes. The region, NGC 2264 -C, was a 
target of many studies to explore its gas dynamics and estimate some of its physical properties such as density and size (e.g. \citealt{per06, per07, mau09}). 
The results of those dynamical studies showed that the material in the region is in a pre-collapse process toward its center which is located close to the core CMM3. Maps of NGC 2264 revealed that outflows may be associated with the core CMM3, but neither of these studies found that the core itself is collapsing. Moreover, none of the previous studies looked into modeling the chemistry neither of the whole region nor of one of its mm-cores.

The clump CMM3 was recently an object of attraction of millimeter observations to understand its chemical content and to derive the physical conditions that may govern its chemistry \citep{sar11, wat15}. \citet{sar11} conducted high resolution observations of molecular outflows in the CMM3 region using the carbon monoxide, CO, and methanol, CH$_3$OH, molecular lines. 
The results suggested that CMM3 is hosting a young massive protostar in its early evolutionary stages. 
Another supporting evidence that CMM3 has an ongoing star formation activity lies in the detection of methyl formate, HCOOCH$_3$, emission lines from this core using interferometric observations \citep{sak07}. The authors showed that the detection of HCOOCH$_3$ emission has an extended distribution similar to that obtained for several hot core regions (e.g. \citealt{beu05}), and its origin is the evaporation from dust grains due to the heat of embedded young stellar object (YSO). Most recently, \citet{wat15} conducted mm-spectral line survey toward NGC 2264 CMM3 in the 0.8, 3, and 4 mm bands in which they detected a total of 339 emission lines. 
They detected strong lines of deuterated species as well as many carbon-chain molecules such as cyanobutadiyne, HC$_5$N, butadiynyl, C$_4$H, and the radical C$_2$S. Weak emission lines of complex organic molecules such as dimethyl ether, CH$_3$OCH$_3$, and methyl formate, HCOOCH$_3$ were also detected. Detection of emission lines of CH$_3$OH with high upper state energies indicates the existence of a hot core. \citet{wat15} concluded that CMM3 shows chemical youth. 

The rich chemistry observed in CMM3 motivated us to introduce the first astrochemical modeling of this clump in which we aim at exploring the core chemistry to understand its chemical evolution, and to obtain an estimate of the core age.

This paper is organized as follows; a detailed description of the chemical models and their initial physical parameters are given in \S\ref{mod}. Model results are presented and discussed in \S\ref{res}. Finally, conclusions are summarised in \S\ref{conc}.

\section{Chemical Models and Initial Physical Parameters}
\label{mod}
The chemical models used in this work are similar to those used in \citet{awad16}, but with physical conditions suitable to the modeled object 
NGC 2264 CMM3, see Table \ref{tab:initial}. The model performs single point computations of the fractional 
abundances of the species, $x$(X), with respect to the total number of H nuclei, n$_{\text H}$, as a function of time at a point of visual extinction, 
A$_v$, of 314 mag calculated self-consistency in the code. This value is inline with the A$_v$ range observed by \citet{war20}. 
According to previous dynamical studies, since the clump CMM3 is not collapsing we can model it as a quiescent core. In modeling these 
static cores, it is widely acceptable to assume uniform density distribution such as the case of modeling hot cores and corinos around massive and 
low-mass star forming regions, respectively 
(e.g. \citealt{viti04, awad10, awad14}). 
In our chemical models, the chemistry happens in a core which is irradiated, from one side, with the standard Draine interstellar radiation field, G$_0$, at a uniform density of 5.4 $\times$ 10$^{6}$ cm$^{-3}$ \citep{per06}, and a kinetic temperature of 15 K. 

Cosmic ray ionisation rate, $\zeta$, is a critical and fundamental quantity for which no direct measurements exist in both diffuse and dense clouds. This rate is poorly constrained by observations because low-energy cosmic rays (CR), that are key sources of molecular ionisation in interstellar clouds, are easily deflected by magnetic field in the interstellar medium (ISM) and solar system \citep{pad09, tiel13}. 
CR ionisation is the key process in initiating the gas-phase chemical network in astrophysical regions via the formation of protonated molecular hydrogen, H$_3^+$, via ``H$_2$ + CR $\longrightarrow$ H$_2^+$ + H$_2$ $\longrightarrow$ H$_3^+$ + H'' (e.g. \citealt{hogn25, tiel13} and references therein for a review). 
For this reason, H$_3^+$ can give a direct measurement to the rate of CR ionisation, $\zeta$, in a given region. To our knowledge, there is no estimates for the CR ionisation rate in CMM3, $\zeta_{{\text {CMM3}}}$. 
In dense cores with column densities (10$^{22} <$ N(H$_2$) cm$^{-2} <$ 10$^{25}$), \citet{pad09} found that the CR ionisation rate, $\zeta$, decreases by a power-law approximated by the formula in Eq. \ref{eq:0}
\begin{eqnarray}
\label{eq:0}
 \zeta ~\sim ~\zeta_{0,k} ~~\Bigg[\frac {\text {N(H$_2$)}} {10^{20} \text {cm}^{-2}}\Bigg]^{-a}
\end{eqnarray}
where the $\zeta_{0,k}$ and $a$ are the fitting constants. Considering the CR distribution of \citet{web98}, which take into account the low-energy component, Padovani et al. found these constants to be 2.0 $\times$ 10$^{-17}$ s$^{-1}$ and 0.021, respectively. Substituting by these values and N(H$_2$) = 5.7 $\times$ 10$^{23}$ cm$^{-2}$, for CMM3 from \citet{per06}, into the expression in Eq. \ref{eq:0}, we obtain $\zeta_{{\text {CMM3}}} =$ 1.67 $\times$ 10$^{-17}$ s$^{-1}$ which we adopt for our reference model. This rate is 1.3 times the standard rate in the ISM ($\zeta_{{\text {ISM}}} =$ 1.3 $\times$ 10$^{-17}$ s$^{-1}$; \citealt{lep92}), and it is in agreement with previous studies \citep{vant20}. However, since $\zeta$ was recorded to have wide range (0.8 $\times$ 10$^{-17} < \zeta ({\text {in s}}^{-1}) <$ 2 $\times$ 10$^{-16}$) in different dense cores around massive protostellar objects (e.g. \citealt{vant20, doty02, moro14, kaz15}), we performed some calculations with higher rates, $\zeta_{\text{CMM3}} \sim$ (5 - 10) $\times \zeta_{\text{ISM}}$, as described in the context below.

Chemical reactions take place both in gas-phase and on grain surfaces. Mantle species are returned to the gas-phase by non-thermal desorption mechanisms: (i) H$_2$ formation on grains, (ii) direct CR heating, and (iii) CR induced photoreactions. The rates of these processes have been adopted from \citet{robj07}, and they are calculated self-consistency in the code at each time step. Due to the low temperature of the core, 15 K, thermal evaporation is suppressed.

The gas-phase chemistry is based on the latest release of the astrochemical UDfA\footnote{Also known as UMIST; 
http://udfa.ajmarkwick.net/} database, UDfA2012 \citep{mce013}. Desorption energies of molecules were updated from KIDA database\footnote{KIDA: KInetic Database for Astrochemistry; http://kida.obs.u-bordeaux1.fr}, when needed. We also updated the gas-phase network to include the formation reactions of HCOOCH$_3$ and CH$_3$OCH$_3$ (shown below) as recently investigated by \citet{bal15}
\begin{align*}
&{\text {CH$_3$OH + OH}} \longrightarrow  {\text { CH$_3$O + H$_2$O}} &\quad ~\alpha = 3.0 \times 10^{-10}, ~\beta = -1, ~\gamma = 0 \\
&{\text {CH$_3$O + CH$_3$}} \longrightarrow  {\text { CH$_3$OCH$_3$ + h$\nu$}} &\quad \alpha = 3.0 \times 10^{-10}, ~\beta = 0, ~\gamma = 0 \\
&{\text {CH$_3$OCH$_3$ + Cl}} \longrightarrow  {\text { CH$_3$OCH$_2$ + HCl}} &\quad \alpha = 2.0 \times 10^{-10}, ~\beta = 0, ~\gamma = 0 \\
&{\text {CH$_3$OCH$_3$ + F}} \longrightarrow  {\text { CH$_3$OCH$_2$ + HF}} &\quad \alpha = 2.0 \times 10^{-10}, ~\beta = 0, ~\gamma = 0 \\
&{\text {CH$_3$OCH$_2$ + O}}  \longrightarrow {\text { HCOOCH$_3$ + H}}  &\quad \alpha = 2.0 \times 10^{-10}, ~\beta = 0, ~\gamma = 0 
\end{align*}
The surface network includes simple hydrogenation of species in addition to the mantle formation of CH$_3$CN, because its gas-phase formation is not efficient (e.g. \citealt{tiel82, gar08}), and HCOOCH$_3$ which is recently studied by \citet{occ11}. The prefix `m' in the pathways below indicates mantle species
\begin{align*}
\label{eq:01}
&{\text {mCH$_4$}} + {\text {mHCN}} \longrightarrow {\text {mCH$_{3}$CN + H$_2$}} \qquad \alpha = 1, ~\beta = 0, ~\gamma = 0 \\
&{\text {mCH$_3$OH}} + {\text {mCO}} \longrightarrow {\text {mHCOOCH$_{3}$}} \qquad \alpha = 6.2 \times 10^{-18},~ \beta = 0,~ \gamma = 0 \\
\end{align*}
The rates of reactions are calculated based on the mathematical formulae described in UMIST database releases (e.g. \citealt{mil97,let20,woo07,mce013}) that are:
$$k = 
\begin{cases} 
\mbox{a) Two-body reaction}\\
\alpha ~ (T/300)^{\beta} ~ exp(-\gamma/T)~~ \mbox{cm$^3$ s$^{-1}$} \\ \\
\mbox{b) CR ionisation reaction, for $\zeta = \zeta_{\text{ISM}}$}\\
\alpha \quad\quad \mbox{s$^{-1}$}
\end{cases} 
$$%
where $\zeta $ is the CR ionisation rate, and the constants $\alpha$, $\beta$, and $\gamma$ are listed in the astrochemical databases for each reaction. 

The chemical network links 162 species into 1645 chemical reactions both in gas-phase and on grain surfaces. The initial elemental abundances are taken from \citet{asp09} while the physical conditions of the core CMM3 are those obtained observationally by \citet{per06}. Table \ref{tab:initial} lists these initial elemental abundances and the physical parameters of the Reference Model, {\it hereafter} RM, as well as the species of interest in this study. 


In addition to the RM we ran a grid of six other models to explore the influence of varying the core physical parameters: (i) the density, (ii) the depletion, and (iii) the CR ionisation rate, on the molecular abundances of the species in the core. The depletion (or freeze-out) efficiency is determined by the amount of the gas-phase material frozen onto the grains. Our calculations of the depletion efficiency are based on the abundance of CO molecules which are the most abundant mantle species after H$_2$. The depletion efficiency is regulated by varying the sticking probability of gas-phase species onto grain surfaces which is a time-independent factor \citep{raw92}\footnote{For more details about the description of depletion efficiency in chemical models, the reader is referred to \citealt{viti99, viti04, ler08, awad14}.}. Therefore for a given chemical model, the depletion is kept fixed throughout the period of the calculations. 
Table \ref{tab:grid} lists the grid of models we performed and the values of the physical parameters we are investigating in comparison with the RM. In the following context we briefly describe the grid of models we performed. 

Model M1 represents the case of dense core with the density 5 times higher than that of the RM ($n_{\text{H}}$ (M1) = $n_{\text{H}}$ (RM) 
$\times$ 5 = 2.7 $\times$ 10$^{7}$ cm$^{-3}$) while model M2 shows the case of less-dense cores with the density 5 times less than that of RM 
($n_{\text{H}}$ (M2) = $n_{\text{H}}$ (RM) $\div$ 5 = 1.0 $\times$ 10$^{6}$ cm$^{-3}$). Changes in the depletion percentage on grain surfaces are 
simulated by running two models; model M3 in which the depletion efficiency is 15\% lower than that of the RM, which is an arbitrary choice 
following \citet{awad10}, and model M4 where the depletion is set to 0\% to mimic the case of pure gas-phase models in order to test their 
efficiency in supplying the gas with large molecules in the protostellar core following \citet{bal15}. The effect of increasing the CR ionisation 
rate on the core chemistry is investigated in models M5 and M6 where $\zeta_{\text{CMM3}}$ is enhanced by a factor of (5 - 10) $\times \zeta_{\text{ISM}}$. 

\section{Results and Discussion}
\label{res} 
We calculated the fractional abundances ($x$(X) = n(X)/n$_{\text H}$) of a selected set of species includes H$_2$CO, large molecules (a species is 
considered large when it contains 6 or more atoms as defined by \citealt{her09}), and sulphur bearing species to represent part of the identified 
molecules in CMM3 by \citet{wat15}. The results are illusterated in Figures \ref{fig:1}, \ref{fig:2}, and \ref{fig:4} that show trends of the 
time evolution of the molecular abundances in CMM3 under different physical conditions; core density, the amount of depletion, and CR ionisation 
rate, respectively. Moreover, an estimation of the fractional abundance of CS molecule and the core age are also shown and discussed in this 
section.

\subsection{Chemical Trends}
\begin{figure*}
\begin{center} 
\includegraphics[width=18cm]{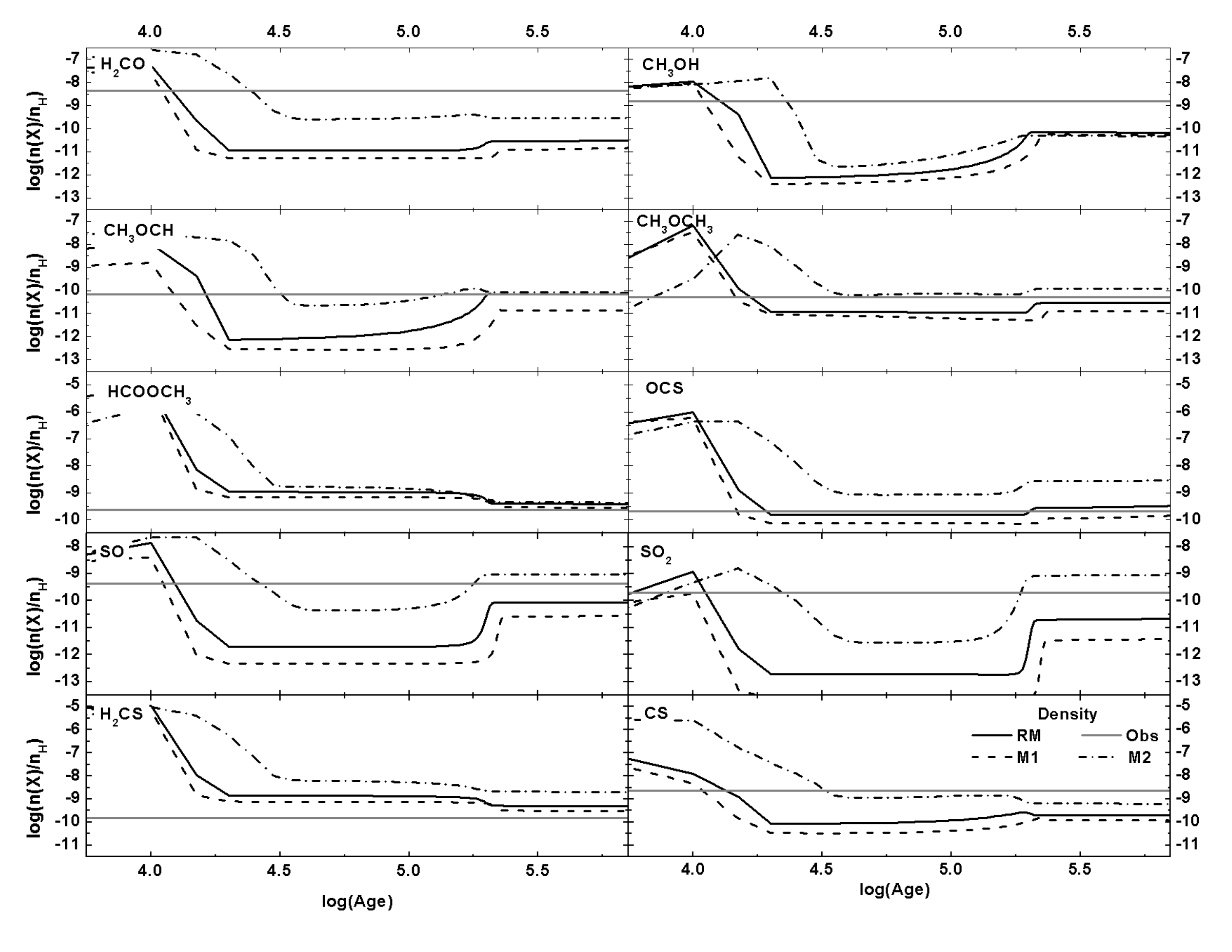} 
\caption {The effect of varying the core density on the time evolution of the calculated molecular abundances. Different curves represent different models: the reference model (RM, n$_{\text H}$ = 5.4 $\times$ 10$^6$ cm$^{-3}$) is the solid line, dense core model (M1, n$_{\text H}$ = 2.7 $\times$ 10$^7$ cm$^{-3}$) is the dashed line, and low-density core model (M2, n$_{\text H}$ = 1.0 $\times$ 10$^6$ cm$^{-3}$) is the dash-dot line. Observed values, as taken from \citealt{wat15}, are represented by solid gray line.}
\label{fig:1}
\end{center}
\end{figure*}

\begin{figure*}
\begin{center} 
\includegraphics[width=18cm]{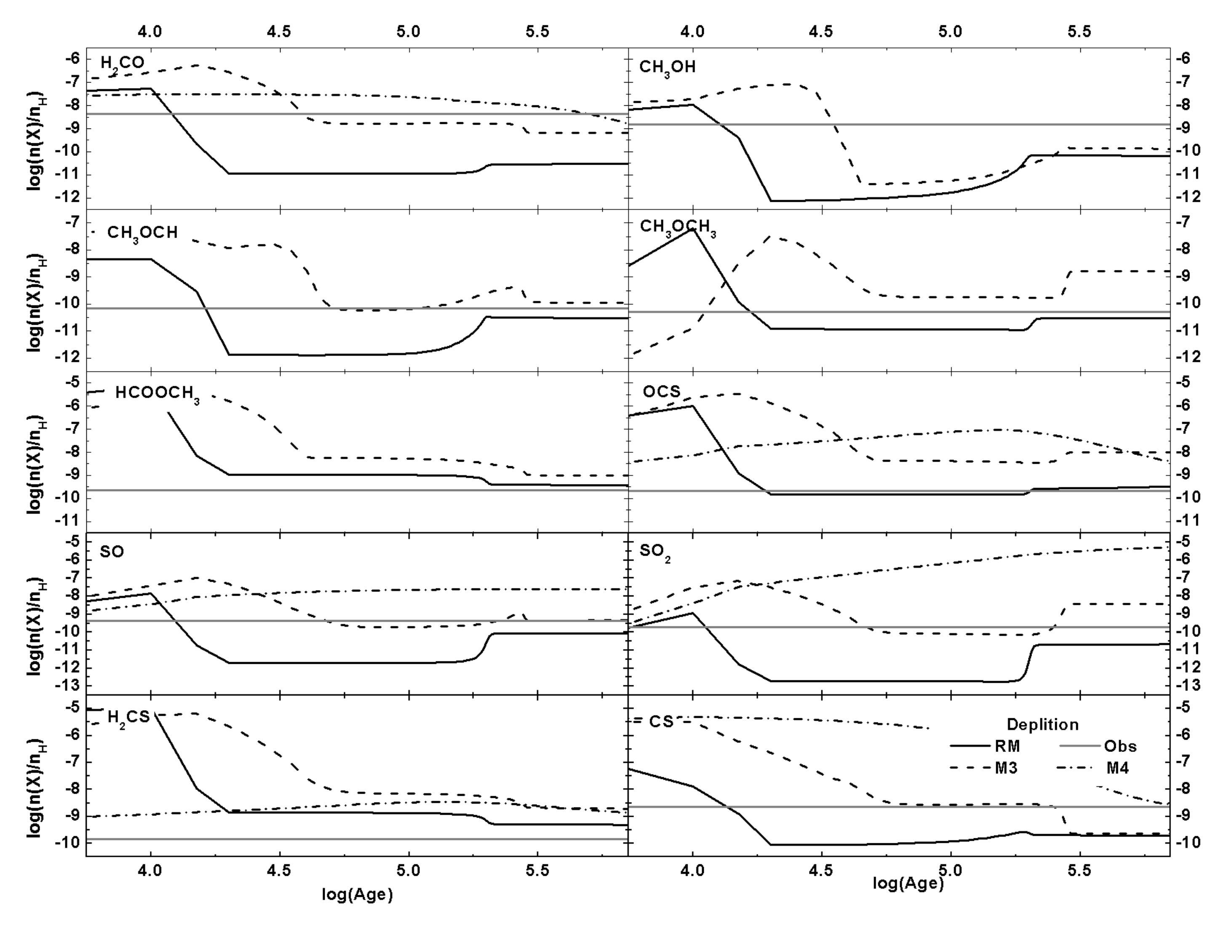} 
\caption {The time evolution of the fractional abundances in cores with different depletion percentage onto grain surfaces. Model M3 (dashed line) 
is 15\% less than the reference model (RM, solid line) and model M4 has 0\% depletion (dash-dot line). Gray straight lines denote observed abundances by \cite{wat15}.}
\label{fig:2}
\end{center}
\end{figure*}

Fig. \ref{fig:1} shows the changes in the core chemistry in models with higher (M1, dash line) and lower (M2, dash-dot line) core densities than 
that in the RM (5.4 $\times$ 10$^6$ cm$^{-3}$, solid line). The effect of varying the depletion of species on their calculated abundances is shown 
in Fig. \ref{fig:2} in which model M3 (dash line) represents partially depleted gas while model M4 (dash-dot line) is the case of non-depleted 
gas. In Fig. \ref{fig:4} we illustrate the calculated molecular abundances in cores where the cosmic ray ionisation rate is 5 times (model M5, dash line) and 10 times (model M6, dash-dot line) higher that the value of the standard rate, $\zeta_{\text{ISM}}$, which are the preference values for thr ionisation rates in massive cores as indicated by observations (e.g \citealt{vant20, doty02}). In all figures observations are represented by solid straight gray line. 

Our results, of all models, show similarities to the evolutionary curves obtained from calculations of the RM, see Figs. \ref{fig:1}, \ref{fig:2}, and \ref{fig:4}. In general, one can notice that fractional abundances have three stages during their time evolution. During early stages of the evolution (t $\le$ 2 $\times$ 10$^4$ yrs), the abundances are very high then they decline to a minimum value before they reach plateau at the second stage at the time interval (2 $\times$ 10$^4 <$ t (yrs) $<$ 3 $\times$ 10$^5$). After that an increase in the abundances is observed before the chemistry reaches another period of steady-state at times later than 3 $\times$ 10$^5$ yrs which is the third phase of the chemical evolution. The three evolutionary phases are explicable in general terms when we look into the chemical analysis.

From the chemical analysis we found that the chemistry that dominates the core, under different physical conditions, shows similarities in the formation and destruction pathways of the studied species. Different curves indicate higher/lower rates of formation and destruction than those calculated for the RM. During the early phase (t $\le$ 2 $\times$ 10$^4$ yrs) of the core chemical evolution, the production of all species is higher than their destruction due to the lack of destructive species. The abundance of these species grows as time passes and when they become enough to affect the yield of a given molecule we see a decline in its abundance ($\sim$ 2 $\times$ 10$^4$ yrs). After that the rates of formation and loss become comparable to each other (within a factor $\le$ 5) causing the obtained plateau during the second phase of evolution at times (2 $\times$ 10$^4 <$ t (yrs) $<$ 2 $\times$ 10$^5$). At the late evolutionary stage, t $\sim$ 3 $\times$ 10$^5$ yrs, the abundances of the molecules begin to increase again due to the consumption of their destructive species. Then, another period of steady-state is reached at times $>$ 3 $\times$ 10$^5$ yrs (see Figs. \ref{fig:1}, \ref{fig:2}, and \ref{fig:4}).

\subsection{Influence of Variation of the Core Physical Parameters}
In this section we discuss the impact of changing three of the core physical parameters on its chemical evolution; namely (1) the core density, (2) depletion onto grains, and (3) the rate of CR ionisation.
\label{trnd}
\subsubsection{The Core Density}
\label{den}
By increasing the core density an increase in the abundances of the species is expected, but this is not what we obtained, see Fig. \ref{fig:1}. 
We found that changes in the core density have minor effects in denser cores (model M1, dash line) by reducing all molecular abundances while in less-dense cores (model M2, dash-dot line) the abundances are enhanced. These increases in the abundances, during the second evolutionary stage, in model M2 are minor ($\le$ factor of 5) for CH$_3$OH, CH$_3$OCH$_3$, HCOOCH$_3$, carbonyl sulphide, OCS, and thioformaldehyde, H$_2$CS, but they are more pronounced ($\ge$ 10 times the abundances in the RM) for formaldehyde, H$_2$CO, acetaldehyde, CH$_3$OCH, sulphur monoxide, SO, sulphur dioxide, SO$_2$, and carbon monosulphide, CS. From the chemical analysis we found that the reason for obtaining this unexpected result is due to either the change in the efficiency of the formation and destruction routes as a result of the variability of the abundance of the parent molecules or due to different chemistry in the mimicked environment. 

\begin{figure*}
\begin{center} 
\includegraphics[width=10cm]{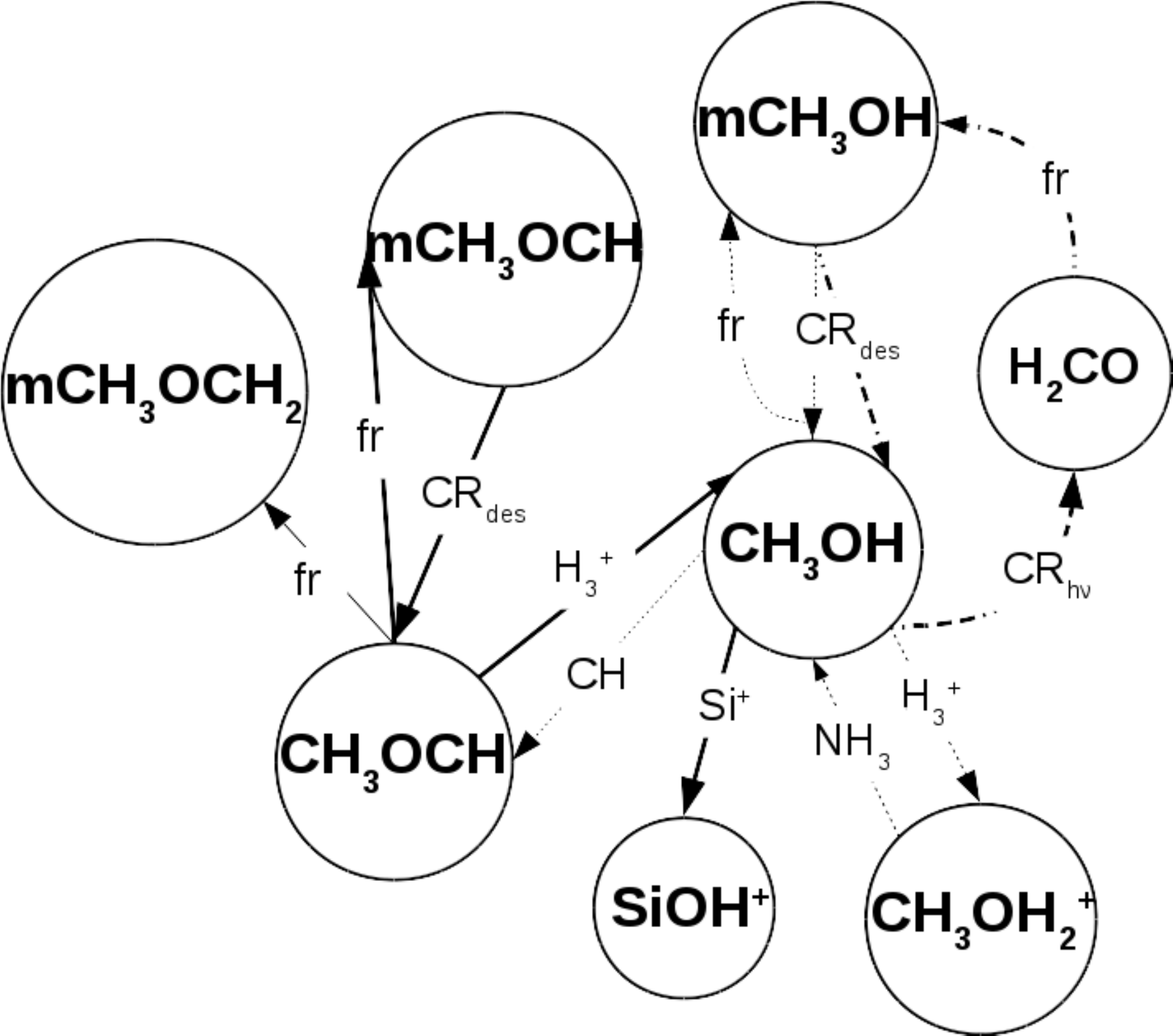} 
\caption {The chemical network for CH$_3$OH and CH$_3$OCH molecules as revealed by the chemical analysis of the RM. The network shows the main formation and destruction pathways of the species. The prefix `m' indicates mantle species. Different line styles for the arrows represents different situations: short dashed arrows are the pathways that become weak with time while thick solid lines are the dominant routes until late times of evolution $>$ 10$^5$ years. Dash-dotted arrows represent extra formation routes of CH$_3$OH that are minor in the RM, during early times $<$ 2 $\times$ 10$^4$ yrs, but dominate the chemistry in less dense cores (model M2) at the same time interval; see the text in \S \ref{den}.} 
\label{fig:3}
\end{center}
\end{figure*}

\begin{figure*}
\begin{center} 
\includegraphics[width=18cm]{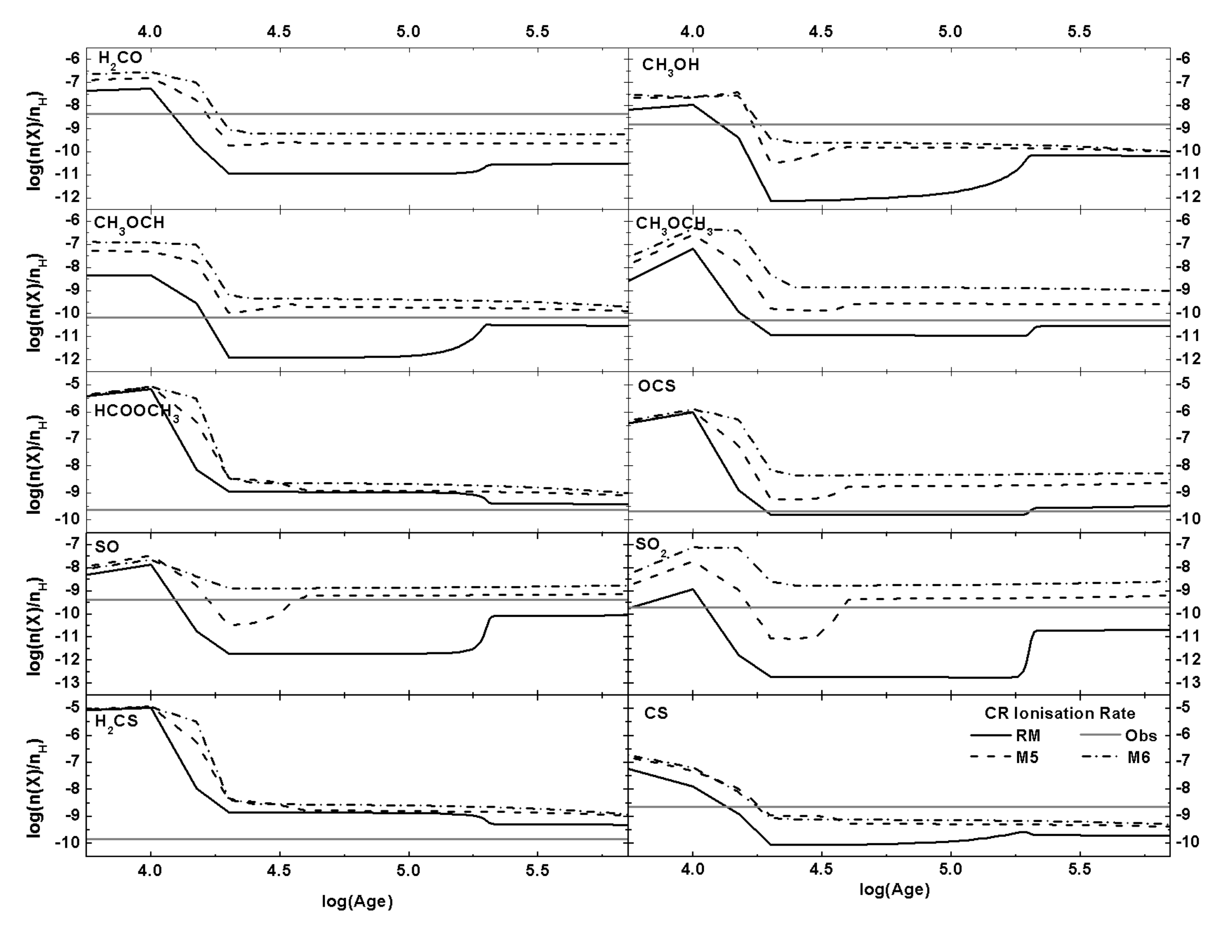} 
\caption {The impact of increasing the CR ionisation rates, $\zeta$, on the calculated molecular fractional abundances in CMM3 compared to our RM (solid curves) and observations (straight gray lines) taken from \citep{wat15}. Dash lines are the calculations in cores with $\zeta = 5\zeta_{\text{ISM}}$ (model M5) while dash-dot lines represent model M6 in which $\zeta = 10\zeta_{\text{ISM}}$.}
\label{fig:4}
\end{center}
\end{figure*}

To explain this finding, we will discuss the detailed analysis of methanol, CH$_3$OH, as an example. A sketch of the analysis of the species in the RM is illustrated in Fig. \ref{fig:3} showing the dominant routes of formation and destruction (solid arrows), the routes that become weak with time (dash arrows), and an extra route that dominates the chemistry of CH$_3$OH in model M2 (dash-dot arrows). The formation of CH$_3$OH in the RM is mainly due to two routes; the first route is via the gas-phase reaction `CH$_3$OH$_2^+$ + NH$_{3} \rightarrow$ CH$_3$OH', the second is due to the evaporation of mantle CH$_3$OH through non-thermal induced UV photodesorption `mCH$_3$OH + CR$_{des} \rightarrow$ CH$_3$OH' (where the prefix `m' referes to mantle species, see short dash arrows in Fig. \ref{fig:3}). At times (t $>$ 5 $\times$ 10$^4$ yrs), the gas phase route becomes inefficient, and the formation of methanol is dominated by another gas-phase reaction `H$_3^+$ + CH$_3$OCH' which is indicated by the solid arrow in Fig. \ref{fig:3}. This reaction dominates the chemistry of CH$_3$OH due to the relatively high abundance of the reactants (parent molecules) at any time later than 5 $\times$ 10$^4$ yrs. Methanol molecules are destroyed by CH and Si$^+$ to form CH$_3$OCH and SiOH$^+$, respectively. The latter dominates the destruction routes at times t $>$ 5 $\times$ 10$^4$ yrs. 

The same formation and destruction pathways are found to control the chemistry of CH$_3$OH in models M1 (denser cores) and M2 (less-dense cores) with comparable formation and destruction rates in model M1 to those of the RM (factor of 2). This finding may explain why the abundance of CH$_3$OH (and other species) in model M1 is slightly less but comparable to that computed in the RM (see Fig. \ref{fig:1}). The abundance of CH$_3$OH, in model M2, is initially enhanced due to its active formation route through surface hydrogenation of H$_2$CO (dash-dot line in Fig. \ref{fig:3}) for longer times comparing to the RM. This route maintain the higher abundance of gaseous CH$_3$OH in less-dense cores for longer times before it declines to converge to the values obtained in the RM around 10$^5$ yrs. 

CH$_3$OH is mainly destroyed via Si$^+$ which starts to decrease in abundance at times $\sim$ 2 $\times$ 10$^5$ yrs. This decrease allows CH$_3$OH to stay longer in the gas causing its abundance to re-increase until it saturates at times $>$ 5 $\times$ 10$^5$ yrs. After that time, the chemical analysis showed that the main pathways as well as the net rates of formation and destruction of CH$_3$OH in all the three models become close. This leads to the observed conversion of the abundances at late times. The discussed scenario for methanol is applicable, in general terms, for the other species in question. 
\subsubsection{The Depletion Efficiency}
\label{gr}
The role of grain surfaces as factories of various interstellar molecules is now well-established (e.g. \citealt{tiel13,linn15}, and references therein). 
Therefore, we expect that decreasing the amount of species freezes onto grain surfaces could cause a decrease in the abundances of their daughter molecules both in gas-phase and on grain mantles.
The sensitivity of molecular abundances to variations in the depletion percentage on grains is illustrated in Fig. \ref{fig:2}. The impact was studied by running two models one with a decreased freeze-out efficiency by 15\% of the RM value following \citet{awad10} (model M3, dashed lines), and another model in which the freeze onto grain surfaces is set to zero to mimic pure gas-phase models (model M4, dash-dotted line). The results showed that the abundances of all selected molecules are affected by changing the depletion efficiency. Sulphur bearing species show an increase in their abundances in both models M3 and M4 while abundances of large molecules are enhanced in model M3 and become invisible in model M4. The latter result emphasises the role of surface reactions in the formation of large molecules (e.g. \citealt{tiel82, hasg92, gar08}). 

The chemical analysis unveiled that the abundance of gaseous species is enriched through their non-thermal desorption via CR induced photodesorption from grain mantles, and when the species are frozen back onto grain surfaces their gaseous abundance are reduced. 
In model M3 and during early stages of the evolution, t $<$ 10$^4$ yrs, the species are produced via few gas-phase channels in addition to their induced photodesorption from grain surfaces. This leads to an increase of their abundances. Moreover, the species are destroyed through a series of gas-phase reactions that have a very small net loss rates $<$ 10$^{-22}$, and hence do not strongly affect the molecular abundances. 
We, also, have noticed that the formation and destruction channels as well as their rates in models M3 and the RM become closer 
at later stages of the evolution (t $>$ 5 $\times$ 10$^5$ yrs) which causes the conversion of the results.

In order to clarify the above discussion on depletion, we discuss the case of CH$_3$OCH. The production of CH$_3$OCH in the RM is via its induced photodesorption from grain surfaces and the gas-phase reaction `CH + CH$_3$OH'. The abundance of the molecule is decreased by its depletion onto grains to either form icy CH$_3$OCH or hydrogenate into CH$_3$OCH$_2$, see Fig. \ref{fig:3}. In model M3, from the chemical analysis we found that the formation of CH$_3$OCH is governed by the gas-phase reaction `CH + CH$_3$OH' because, at any time, the abundance of CH$_3$OH is higher in model M3 than in the RM by 2 orders of magnitudes. The reason for this is that the formation of methanol in model M3 occurs by the reactions `CH$_3$OCH + H$_3^+$' and `NH$_3$ + CH$_3$OH$_2^+$', in which the latter does not weaken with time as the case in the RM, and its destruction by Si$^+$ ions is minor. 
This situation keeps the abundance of methanol higher in model M3 than in the RM and hence increases the yield of CH$_3$OCH. 

When depletion become zero, in model M4, large molecules become undetectable ($x$(X)$<$ 10$^{-13}$ which is the detection limit). This is in agreement with previous models in that large molecules are mainly formed on grain surfaces (e.g. \citealt{tiel82, hasg92, gar08, occ11}). Given the fact that in model M4 the gas is not enriched by any large molecules from grain surfaces highlights their role as molecule suppliers even with small amounts as the case of model M3. The analysis showed that the loss rates of large molecules (species $>$ 6 atoms) in model M4 is higher (up to 50 times) than they are in the RM and that is becuase of the higher abundances of H$_3^+$ and HCO$^+$ ions (by few orders of magnitude) that are the main destroyers of large molecules in the gas. For this reason, gas-phase reactions alone can not reproduce the observed abundances of large molecules in our models. 

On the other hand, calculations of the abundances of sulphur bearing species in model M4 is overestimated up to 5 orders of magnitudes, e.g. the case of SO$_2$, comparing to fully depleted gas in the RM and observations (see Fig. \ref{fig:2}). This result can be directly explained in light of the fact that in dense regions ($n_{\text H} >$ 10$^4$ cm$^{-3}$) sulphur bearing species are believed to be depleted onto grain surfaces \citep{cas94, wak04}. In our study, the chemical analysis of model M4 for these species showed that omitting the depletion of sulphur allows the gas-phase routes of the species to proceed with higher net rates of formation range between 3 - 10 times those of the net rates of destruction at any time. These rates are also higher than those computed in the RM up to 300 times. 
For instance, the molecule SO is destroyed in the RM due to its depletion onto grain surfaces and is increased in the gas-phase due to its non-thermal desorption events and the reaction `OH + S'. When the depletion is set to zero in model M4, we found that the molecule is formed by the latter pathway in addition to `C + SO$_2$' until core age of 3.5 $\times$ 10$^4$ yrs. We also found that the net formation rates are 300 times higher than those in the RM, but the net loss rates are comparable in both models. At ages (t $>$ 4 $\times$ 10$^4$ yrs), the formation through `C + SO$_2$' becomes inefficient and the formation occurs mainly via `OH + S'. Moreover, the net rates of formation and destruction are very close at that time which may explain the steady state that the SO molecules experience at times $\ge$ 30,000 years.

From the above discussion, in \S\ref{den} and \S\ref{gr}, we can conclude that sulphur bearing species, apart from OCS and H$_2$CS, are better represented in protostellar cores with either fully depleted gas and lower core density (1 $\times$10$^{6}$ cm$^{-3}$) or partially depleted gas and the standard core density (5.4$\times$10$^{6}$ cm$^{-3}$). 
The abundance of OCS better matches observations (gray lines in Figs. \ref{fig:1} and \ref{fig:2}) in models with dense ($>$ 10$^6$ cm$^{-3}$) and fully depleted gas; i.e the RM and M1. This may indicate that the origin of OCS is icy grain mantles. This finding is inline with the detection of solidified OSC in AFGL 989 and Mon R2 IRS 2 by \citet{pal97}. 

\subsubsection{The Cosmic Ray Ionisation Rate}
\label{cr}
Cosmic ray ionisation is an important process in the ISM because it is the key process in the production of H$_3^+$ which is the initiator of the gas-phase chemistry in astrophysical regions. Hence, the rate of CR ionisation is a fundamental parameter in astrochemical modeling \citep{tiel13}. 
Several studies showed that the CR ionisation rate, $\zeta$, in massive star-forming regions are much higher than the standard value, $\zeta_{\text {ISM}}$ = 1.3 $\times$ 10$^{-17}$ s$^{-1}$, by 5 - 500 times with a preference values ($\sim$ 5 - 10)$\times \zeta_{\text {ISM}}$ (e.g. \citealt{vant20, doty02, moro14, kaz15}). In order to investigate the impact of increasing $\zeta$ on the chemistry of CMM3, we run models with higher ionisation rates. Model M5 at $\zeta =$ 5 $\zeta_{\text {ISM}}$ and model M6 at $\zeta =$ 10 $\zeta_{\text {ISM}}$ that are denoted by dash and dash-dot lines, respectively, in Fig. \ref{fig:4}. The solid black and gray lines represent the calculations of the RM and the observations by \citet{wat15}, respectively. 

In general, we noticed that the evolutionary trends of all the species show similarities to the trend obtained in the RM which implies that the chemistry in action does not change but changes occur in the efficiency of the chemical pathways as a results of variations in the physical parameters of the core (as previously discussed). Increasing the CR ionisation rate enhances the abundances of the studied species (see Fig. \ref{fig:4}). The least affected molecules by variations in $\zeta$ are HCOOCH$_3$ and H$_2$CS (increased $\le$ 5 times) while the most influenced species (increased $\ge$ 200 times) are CH$_3$OH, CH$_3$OCH, SO, and SO$_2$. The chemical analysis revealed that during early times, t $<$ 2 $\times$ 10$^4$ yrs, the net rate of formation of the species increased up to 100 times their values in the RM while the loss rates are 10 - 100 times lower than these formation rates. This leads to the observed increase of the molecular abundances at this early stage. Moreover, by increasing the CR ionisation rate ions and radicals (that are common parent molecules) become more abundant in the gas. As a consequence, the yield of their daughter molecules increases comparing to the RM. 

From Fig. \ref{fig:4} we notice that the results of model M6 with $\zeta =$ 10 $\zeta_{\text {ISM}}$ are, generally, overestimating the abunances of the species while the results of model M5 with $\zeta =$ 5 $\zeta_{\text {ISM}}$ are in better agreement with observations (gray lines in the figure). Therefore, we may expect the CR ionisation rate in the core CMM3 to be not more than 5 times the standard value $\zeta_{\text {ISM}}$. This result is inline with observations of cores around massive protostellar objects \citep{cas98, vant20, doty02}. 

\subsection{The abundance of CS molecules}
\label{abund}
\citet{wat15} identified around 36 molecules in CMM3 core. In their Table 8, the observed values of the species are given in column densities (cm$^{-2}$) while our models compute fractional abundances with respect to the total H in all forms ($n_{\text H}$). Therefore, we converted all the observed values into fractional abundances assuming the column density of H$_2$ in CMM3 is equal to 5.7$\times$10$^{23}$ cm$^{-2}$ as obtained by \citet{per06}. 

Since the hydrogen column density at one magnitude of visual extinction (1 A$_v$) is 1.6 $\times$ 10$^{21}$ cm$^{-2}$, then the total column density of hydrogen at a particular point of visual extinction A$_v$ can be expressed as `1.6 $\times$ 10$^{21}$ * A$_v$' \citep{snow06}. 
Hence, for any species, Y, at a point of visual extinction A$_v$ magnitudes, its calculated fractional abundance, $x$(Y), in our models can be converted into 
column density, N(Y), using Eq. \ref{eq:1}
\begin{eqnarray}
\label{eq:1}
 {\text N}({\text Y}) = 
x({\text Y}) * 1.6 \times 10^{21} * {\text A_v}   
\end{eqnarray}

The detection of C$^{33}$S and C$^{34}$S was reported in Tables 6 and 8 in \citet{wat15}. 
In the local ISM and the solar system, the isotope ratio of $^{33}$S and $^{34}$S with respect to the normal isotope $^{32}$S (simply CS) was calculated to be 22 \citep{wil94}. Given this ratio and the observed column density of C$^{34}$S in CMM3 (N(C$^{34}$S) = 1.2$\times$10$^{14}$ cm$^{-2}$), we calculated N(CS) by applying the expression in Eq. \ref{eq:2}
\begin{eqnarray}
\label{eq:2}
\text N({\text {CS}}) = (^{32}{\text S}/^{34}{\text S}) * {\text N}({\text C}^{34}{\text S})   
\end{eqnarray}
which gives a value; N(CS) = 2.6 $\times$ 10$^{15}$ cm$^{-2}$ from which we calculated the fractional abundance to be, $x$(CS) = 2.3 $\times$ 10$^{-9}$. This value is used in comparing our theoretical calculations with observations as listed in Table \ref{tab:comp}.

\subsection{CMM3 Age Estimation}
\label{comp}
We selected our set of species to represent part of the different species identified observationally in the massive protostellar core NGC 2264 CMM3 by \citet{wat15}. Observations of this core concluded that it shows characteristics of chemical youth (e.g. \citealt{sar11, wat15}). Although \citet{lada93} estimated the age of the whole massive cluster forming region NGC 2264 to be $\sim$ 5$\times$10$^6$ yrs, none of the following studies were able to estimate, quantitatively, the age of the core CMM3 itself.

Results of the chemical models were able to give an estimation to the age of CMM3 core by comparing them to the observations of \citet{wat15} (straight gray line in all figures). Calculations of the RM (solid curve in all figures) best fit observations at times $\le$ 20,000 yrs. In the case of methyl formate, HCOOCH$_3$, the RM calculations are in good agreement with observations by \citet{sak07} during early time stages (t $<$ 2 $\times$ 10$^4$ yrs) while it is in better agreement with observations of \citet{wat15} at later times, t $>$ 3.2 $\times$ 10$^5$ yrs. Models with variations in their physical parameters are often best fit observations during later times stages, (2 - 3) $\times$ 10$^4$ yrs, comparing to those of the RM (see Figs. 1, 2, and 4). Table \ref{tab:comp} list the comparison between observations of CMM3 and the RM calculations, and it also gives the time of best fit. 

Variations in the core density, Fig. \ref{fig:1}, influence the time of best fit of the calculated molecular abundances to observations. The chemical age of the core is found to be inversely proportional to the core density. Calculated abundances in models of dense cores better fit observations at early times ($\sim$ 10,000 yrs) while results of less-dense cores are in agreement with observations at slightly later times ($\sim$ 32,000 yrs). The plausible reason for this result is that for less-dense cores, the chemistry may take longer time to build enough amounts of reactants that will , in turn, enrich the material with their daughter molecules when their abundances are high enough.

Changes in the depletion efficiency of the gas by 15\%, Fig. \ref{fig:2}, affect the calculated abundances of the species in the core leading to an agreement with observations for few species at later times ($\sim$ 56,000 yrs) comparing to the times of best fit in the RM. S-bearing species are observable under low- density and depletion, our results showed that the abundance of SO, SO$_2$, and CS best fit observations in models with either low density at t $\le$ 32,000 yrs or partial depletion at t $\sim$ 5.6 $\times$ 10$^4$ yrs. On the other hand, pure gas-phase models can not reproduce observations for neither large molecules nor S-bearing speccies (see Fig. \ref{fig:2}). 

Moreover, enhancing the CR ionisation rate, $\zeta$, in the studied core, Fig. \ref{fig:4}, leads to an increase of the fractional abundances of all species that , in turn, results in a better agreement with observations than that obtained for the RM. However, in general, when $\zeta \sim$ 10 times the standard value, our models overestimate the abundances of the species. Models with $\zeta \sim$ 5 $\zeta_{\text{ISM}}$ best fit observations at times around 17,000 yrs, that is comparable to the time of best fit of the RM by a factor of $\sim$ 2. 

To sum up, we conclude that massive protostellar cores with density range (1 - 20) $\times$ 10$^6$ cm$^{-3}$ and have CR ionisation rates between 1.3 and 5 $\zeta_{\text{ISM}}$ may have age range 1 $\times$ 10$^4$ $\le$ age (yrs) $\le$ 5.6 $\times$ 10$^4$. This age is comparable to the estimated age of massive protostellar cores \citep{doty02}, and young cores that host Class 0 objects \citep{and93, and20}. Thus, our theoretical modeling supports previous suggestions that CMM3 is chemically young massive core that may host Class 0 object \citep{sar11, wat15}. 

\section{Conclusions}
\label{conc} 
This work represents the first astrochemical modeling of the protostellar core CMM3 in the massive cluster-forming region NGC 2264. We studied the impact of varying three of the core physical conditions: the gas density, depletion of species, and the CR ionisation rate, on the chemical evolution of the molecular fractional abundances (n(X)/n$_{\text H}$). The results showed that both large and sulphur-bearing molecules response to variations in these parameters, but their sensitivity to these changes is different. From the discussion of the results, we can conclude the following:
\begin{enumerate}
\item [{\bf 1-}] Cores with densities (5 - 10) $\times$ 10$^6$ cm$^{-3}$ are often in better agreement with observations than less dense ($<$ 5 $\times$ 10$^6$ cm$^{-3}$) cores. This may set a critical densities to other cores with similar chemical contents but unknown densities. 

\item [{\bf 2-}] Models with full depletion and high density are able to reproduce the observed abundances of large molecules and OCS due to their grain mantle origin. Models with either partial depletion or low density are able to reproduce the observed abundances of SO, SO$_2$, and CS at times $\ge$ 2 - 3 $\times$ 10$^4$ years. This indicates that these molecules can be effictively formed through gas-phase routes. 

\item [{\bf 3-}] Pure gas-phase models are not able to reproduce the observed abundances of neither large molecules nor sulphur bearing species which highlightes the importance of depletion and, hence, surface chemistry in protostellar cores as it is important in hot cores. 

\item [{\bf 4-}] Although we estimated the CR ionisation rate in CMM3 to be $\zeta_{\text{CMM3}} =$ 1.67 $\times$ 10$^{-17}$ s$^{-1}$ which is 1.3 $\zeta_{\text{ISM}}$, higher ionisation rates than $\zeta_{\text{CMM3}}$ up to 5 times can also best fit observations during early stages. Therefore, the CR ionisation rate in CMM3 may range between (1.3 - 5) $\times$ $\zeta_{\text{ISM}}$. This range of ionisation rate is inline with previous observational estimates in massive cores.

\item [{\bf 5-}] The calculated abundances of the species in the RM are in good agreement with observations during the early stages of the core chemical evolution $\sim$ 1 - 2 $\times$ 10$^4$ yrs. Moreover, abundances of molecules in protostellar cores either possesses density range of (1 - 20) $\times$ 10$^6$ cm$^{-3}$ or CR ionisation rate up to 5 times $\zeta_{\text{ISM}}$ best fit observations during times 2 $\times$ 10$^4 <$ t (yrs) $<$ 5 $\times$ 10$^4$. These times are in agreement with the estimated age of Class 0 protostellar cores \citep{and93, and20}. Therefore, we conc;ude that CMM3 is a young massive protostellar core that may host Class 0 object and its age is in the range (1 - 5) $\times$ 10$^4$ yrs.

\end{enumerate}
\section*{Acknowledgments}
{\bf ZA} would like to thank {\bf Dr. Yoshimasa Watanabe} for the time he granted her to explain issues about the observations in \citet{wat15}, 
during their meeting in ISM2016 workshop held in Japan in October 2016, and for the fruitful discussion and comments that improved the original manuscript. 

{\bf The authors} are grateful for the referee for his constructive comments that improved the original manuscript.


\clearpage
\begin{table*}
   \centering
\caption{Initial elemental abundances and physical conditions utilized in this study and the set of species of interest.}
  \label{tab:initial}
    \leavevmode
    \begin{tabular}{llllllll} \hline \hline
\multicolumn{2}{c}{\bf Initial abundances$^{\text a}$} &&\multicolumn{2}{c}{\bf Physical parameters$^{\text b}$} &&\multicolumn{2}{c}{\bf Species of interest}\\ \hline
Helium & 8.50 $\times$ 10$^{-2}$ && Core density (cm$^{-3}$) & 5.4$\times$10$^{6}$ && H$_2$CO & OCS\\
Carbon & 2.69 $\times$ 10$^{-4}$ && Core temperature (K) & 15 && CH$_3$OH & SO\\
Oxygen & 4.90 $\times$ 10$^{-4}$ && Core radius (pc) & 0.04 && CH$_3$OCH & SO$_2$ \\
Fluorine & 3.63 $\times$ 10$^{-8}$&& Core Mass (M$_{\odot}$)&40&& CH$_3$OCH$_3$ & H$_2$CS \\ 
Chlorine & 3.16 $\times$ 10$^{-7}$&& $^{\dag}$Depletion & full && HCOOCH$_3$ & CS \\
Sulphur & 1.318 $\times$ 10$^{-5}$&& $^{\ddag}$CR ionisation rate (s$^{-1}$) & $\zeta_{\text{CMM3}}$  &&  & \\ \hline  \hline
\end{tabular}
\flushleft
$\dag$ By full depletion we mean that, for any given species, more than 90\% of its gaseous form is involved into grain surface reactions. while 
this percentage is 15\% less for partial depletion simulation (model M3) following \citealt{awad10}. Pure gas-phase chemistry is mimicked with 
depletion of 0\% (model M4). \\
$^{\ddag} \zeta_{\text{CMM3}}$ = 1.67 $\times$ 10$^{-17}$ s$^{-1}$, estimated in this work following \citet{pad09} (see \S\ref{intro}).\\
References: (a) \citealt{asp09} , (b) \citealt{per06}
\end{table*}
\begin{table*}
   \centering
\caption{The grid of models we performed, given the main different physical parameters compared to the reference model (RM). Note that 
none of the other parameters stated in Table \ref{tab:initial} have been changed.}
  \label{tab:grid}
    \leavevmode
    \begin{tabular}{llll} \hline \hline
{\bf Model} & {\bf Density} & {\bf $^{\dag}$Depletion} & {\bf $\zeta$}\\
& (cm$^{-3}$) &  & \\ \hline
RM & 5.4$\times$10$^{6}$ & full & $\zeta_{\text{CMM3}}$\\
M1 & 2.7$\times$10$^{7}$ & full & $\zeta_{\text{CMM3}}$\\
M2 & 1.0$\times$10$^{6}$ & full & $\zeta_{\text{CMM3}}$\\
M3 & 5.4$\times$10$^{6}$ & partial & $\zeta_{\text{CMM3}}$\\
M4 & 5.4$\times$10$^{6}$ & none & $\zeta_{\text{CMM3}}$\\ 
M5 & 5.4$\times$10$^{6}$ & full & 5 $\zeta_{\text{ISM}}$\\
M6 & 5.4$\times$10$^{6}$ & full & 10 $\zeta_{\text{ISM}}$\\\hline \hline
\end{tabular}
\flushleft
$\dag$ 
The depletion is partial means that it is 15\% less than the RM following \citealt{awad10}. Pure gas-phase chemistry is mimicked in model M4 with 
depletion of 0\% (none).\\
\end{table*}
\begin{table*}
\centering
\caption{Comparison between our model calculations for the reference model (RM) and observations of NGC 2264 CMM3 from Table 8 in \citealt{wat15}. Time of best fit, in years, is also shown in the last column.}
  \label{tab:comp}
    \leavevmode
    \begin{tabular}{lllll} \hline\hline
{\bf Species} & \multicolumn{2}{c}{\underline{\bf Observations$^{\dag}$}} & {\bf This Work} & {\bf Time} \\ [0.8 ex] 
              &  \multicolumn{1}{c}{\bf N(Y)} &   \multicolumn{1}{c}{\bf $x$(Y)} &    \multicolumn{1}{l}{\bf $x$(Y)} & \multicolumn{1}{l}{\bf yr} \\ \hline
{\bf OCS}&2.4 $\times$ 10$^{14}$  & 2.1 $\times$ 10$^{-10}$ & 1.78 $\times$ 10$^{-10}$ & $\ge$ 1.9 $\times$ 10$^{4}$\\ [0.8 ex]  
{\bf SO} &5.0 $\times$ 10$^{14}$  & 4.4 $\times$ 10$^{-10}$ & 3.98 $\times$ 10$^{-10}$ & 1.26 $\times$ 10$^{4}$\\ [0.8 ex]  
{\bf SO$_2$} &2.2 $\times$ 10$^{14}$  & 1.9 $\times$ 10$^{-10}$ & 1.95 $\times$ 10$^{-10}$ & 1.12 $\times$ 10$^{4}$\\ [0.8 ex]  
{\bf H$_2$CS}& 1.7 $\times$ 10$^{14}$  & 1.5 $\times$ 10$^{-10}$ & $\ge$ 5.31 $\times$ 10$^{-10}$ & all times\\ [0.8 ex]  
{\bf CS} $^{\text \bf b}$& 2.6 $\times$ 10$^{15}$  & 2.3 $\times$ 10$^{-9}$ & 2.34 $\times$ 10$^{-9}$ & 1.38 $\times$ 10$^{4}$\\  \hline
{\bf H$_2$CO} & 5.2 $\times$ 10$^{15}$ & 4.5 $\times$ 10$^{-9}$ & 4.07 $\times$ 10$^{-9}$ & 1.2 $\times$ 10$^{4}$ \\ [0.8 ex] 
{\bf CH$_3$OH}&2.1 $\times$ 10$^{15}$ & 1.5 $\times$ 10$^{-9}$ & 1.38 $\times$ 10$^{-9}$ & 1.32 $\times$ 10$^{4}$\\ [0.8 ex] 
{\bf CH$_3$OCH}&7.7 $\times$ 10$^{13}$  & 6.75 $\times$ 10$^{-11}$ & 7.7 $\times$ 10$^{-11}$ & 1.62 $\times$ 10$^{4}$\\ [0.8 ex]  
{\bf CH$_3$OCH$_3$} &5.8 $\times$ 10$^{13}$  & 5.1 $\times$ 10$^{-11}$ & 4.79 $\times$ 10$^{-11}$ & $\ge$ 1.70 $\times$ 10$^{4}$\\ [0.8 ex]  
{\bf HCOOCH$_3$}& 2.6 $\times$ 10$^{14}$  & 2.3 $\times$ 10$^{-10}$ & 4.37 $\times$ 10$^{-10}$ & $\ge$ 2.19 $\times$ 10$^{5}$\\ [0.8 ex]  
                & 4.0 $\times$ 10$^{15}$ $^{\text \bf a}$ & 3.5 $\times$ 10$^{-9}$ & 1.12 $\times$ 10$^{-9}$ & 2 - 20 $\times$ 10$^{4}$\\ \hline \hline
%
%
%
%
%
%
%
\end{tabular}
\flushleft
$^{\dag}$ Observed column densities are converted into fractional abundances with respect to H, considering N(H$_2$) = 5.7$\times$10$^{23}$ cm$^{-2}$ 
for NGC 2264 CMM3 as determined observationally by \citet{per06}\\
$^a$ Observations are obtained by \citet{sak07} toward the CMM3 core.\\ 
$^b$ The values of observed CS is calculated as described in \S\ref{abund} in the text.\\
\end{table*}



\end{document}